\documentclass[aps,prl,twocolumn,showpacs,superscriptaddress,floatfix,notitlepage]{revtex4-1}

\usepackage{pdfsync}
\usepackage{graphicx}
\usepackage{overpic}
\usepackage{amsmath,amssymb,amsfonts}
\usepackage{bm}
\usepackage{color}
\usepackage{ulem}
\usepackage[breaklinks=true,colorlinks,citecolor=blue,linkcolor=blue,urlcolor=blue]{hyperref}

\begin{document}

\title{A Corbino disk viscometer for 2D quantum electron liquids}

\author{Andrea Tomadin}
\email{andrea.tomadin@sns.it}
\affiliation{NEST, Istituto Nanoscienze-CNR and Scuola Normale Superiore, I-56126 Pisa, Italy}

\author{Giovanni Vignale}
\affiliation{Department of Physics and Astronomy, University of Missouri, Columbia, Missouri 65211, USA}

\author{Marco Polini}
\affiliation{NEST, Istituto Nanoscienze-CNR and Scuola Normale Superiore, I-56126 Pisa, Italy}

\begin{abstract}
The shear viscosity of a variety of strongly interacting quantum fluids, ranging from ultracold atomic Fermi gases to quark-gluon plasmas, can be accurately measured.
On the contrary, no experimental data exist, to the best of our knowledge, on the shear viscosity of two-dimensional quantum electron liquids hosted in a solid-state matrix.
In this Letter we propose a Corbino disk device, which allows a determination of the viscosity of a quantum electron liquid from the dc potential difference that arises between the inner and the outer edge of the disk in response to an oscillating magnetic flux.
\end{abstract}


\maketitle

\noindent {\it Introduction.---}The shear viscosity $\eta$, which describes the diffusion of the average momentum density orthogonally to its direction, is one of the cornerstones of the hydrodynamic theory of fluids~\cite{Landau06, batchelor}. The first estimate of the shear viscosity of a dilute gas as a function of its density and temperature was given by Maxwell in his celebrated article on the ``Dynamical Theory of Gases''.
He found that the shear viscosity of a dilute gas is independent of its density, a counterintuitive result that he felt needed immediate experimental testing~\cite{maxwell}.
Recent years have witnessed a surge of interest in the viscous flow of strongly interacting quantum fluids, for which hydrodynamics provides a powerful non-perturbative description~\cite{abrikosovkhalatnikov}.
Experimentally, the shear viscosity of quantum liquids like $^3{\rm He}$ and $^4{\rm He}$ can be measured by a variety of tools including capillary, rotation, and vibration viscometers~\cite{NPFsreview}.
The shear viscosity of cold atomic gases can be  inferred from measurements of  collective modes or by looking at the expansion of the gas in a deformed trap after the trapping potential is turned off~\cite{cao_science_2011,cao_njp_2011,vogt_prl_2012}. The shear viscosity of quark-gluon plasmas can be extracted from elliptic flow measurements at relativistic heavy-ion colliders~\cite{QGP}.
To the best of our knowledge, however,  no protocols exist for measuring the shear viscosity of two-dimensional (2D) quantum electron liquids (QELs) in solid-state matrices~\cite{MBE,other2DEGs}.
This gap is truly surprising in view of the large body of theoretical work~\cite{shear_viscosity_2DQELs} that has been carried out in connection with the shear viscosity of these systems.  In this paper we try and fill the gap by proposing  a method to measure the  viscosity of electrons in a realistic experimental setup.

\begin{figure}[t]
\includegraphics[width=0.8\columnwidth]{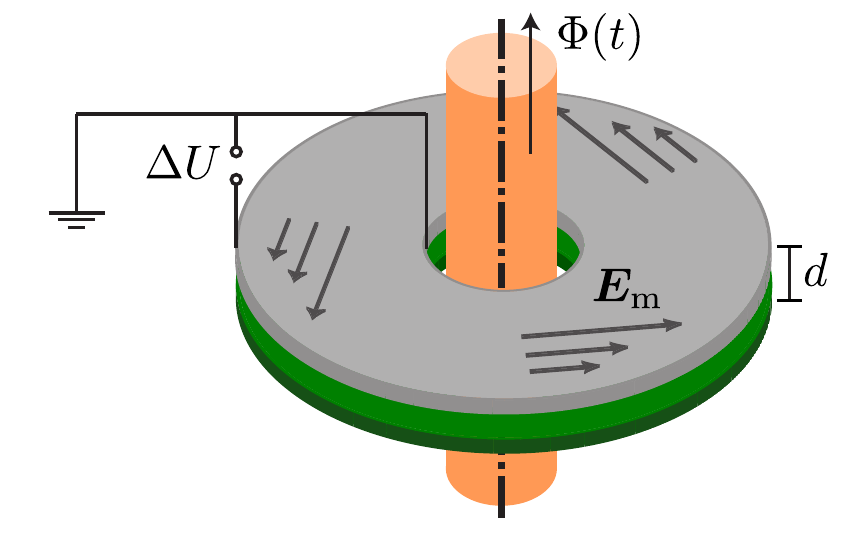}
\caption{(Color online) A viscometer for 2D quantum electron liquids.
The light gray surface represents the 2D electron system, which is shaped into a Corbino disk geometry.
The intermediate green region represents a dielectric layer of thickness $d$, which separates the 2D electron system from a back gate (dark grey region).
The internal hole of the Corbino disk is threaded by an oscillating magnetic flux $\Phi(t) =\Phi_0\cos (\Omega t)$, which induces an azimuthal electric field 
${\bm E}_{\hat{\bm \theta}}$ oscillating at the same frequency $\Omega$.
The magnitude of the azimuthal electric field decreases as one goes from the inner to the outer rim.
The inner rim of the disk is grounded, while the outer rim is free to adjust its voltage.
A dc~electrical potential energy difference $\Delta U$ appears between the inner and the outer rim  in response to the oscillating magnetic flux.
The quantity $\Delta U$ sensitively depends on the shear viscosity of the 2D electron fluid and the frequency $\Omega$ of the magnetic flux. \label{fig:setup}}
\end{figure}

The concept of hydrodynamic viscosity $\eta$~\cite{Landau06, batchelor, abrikosovkhalatnikov} becomes relevant  in a regime of parameters  in which the electron liquid is well described by a quasi-equilibrium distribution function characterized by slowly time-dependent density and drift velocity---the local counterparts of globally conserved particle number and momentum.  In a solid-state device with linear dimension $L$ this is ensured by the following chains of inequalities: $\ell_{\rm ee} \ll L \ll\ell_{\rm p}$, where $\ell_{\rm ee}$ is the mean free path between quasiparticle collisions~\cite{Giuliani_and_Vignale} and the length scale  over which local thermodynamic equilibrium is achieved, while $\ell_{\rm p}$ is the length scale over which electron-impurity and electron-phonon scattering break momentum conservation. It is well known~\cite{dyakonov_prl_1993,govorov_prl_2004,andreev_prl_2011} that the above inequalities are satisfied in highly-pure 2DQELs e.g.~in modulation-doped GaAs/AlGaAs semiconductor heterojunctions~\cite{MBE} for typical electron densities, in the temperature range $5~{\rm K} \lesssim T \lesssim 35~{\rm K}$, and for devices with linear size $10~\mu {\rm m} \lesssim L \lesssim 50~\mu {\rm m}$. In this range of parameters momentum-non-conserving collisions can be safely neglected, while electron-electron interactions establish local equilibrium with a slowly-varying density $n({\bm r},t)$ and drift velocity ${\bm v}({\bm r},t)$. Hydrodynamic electron flow has indeed been experimentally generated and hydrodynamic effects have been measured~\cite{hydrodynamicsexperimental}.  We now describe our method for determining  the viscosity.

\noindent {\it Electrical measurement of the viscosity.---}We consider a 2DQEL shaped in a Corbino disk (CD) geometry---see Fig.~\ref{fig:setup}.
The CD lies in the $z=0$ plane and has an inner radius $r_{\rm in}$ and an outer radius $r_{\rm out}$.
It is separated from a back gate by a dielectric layer of thickness $d$ and dielectric constant $\epsilon$.
An oscillating magnetic flux $\Phi(t) = \Phi_{0} \cos{(\Omega t)}$ oriented along the $\hat{\bm z}$ axis threads the inner hole of the CD 
and induces, by Faraday's law, an azimuthal electric field, which is given by
\begin{equation}\label{eq:emagazim}
{\bm E}_{\hat{\bm \theta}}(r,t) =
-\frac{1}{2 \pi c r} \partial_{t} \Phi(t)  \hat{\bm \theta}~,
\end{equation}
where $c$ is the speed of light and ${\hat{\bm \theta}}$ is the unit vector in the azimuthal direction.
Fluctuations of the circularly-symmetric electron density $n(r,t)$ on the surface of the CD generate a radial electric field of the form
\begin{equation}\label{eq:Eradial}
{\bm E}_{\hat{\bm r}}(r,t) \simeq  \frac{e}{C} \partial_{r} n(r,t) \hat{\bm r}~,
\end{equation}
where $e$ is the absolute value of the electron charge, $C = \epsilon / (4 \pi d)$ is the geometric capacitance per unit area, and ${\hat{\bm r}}$ is the unit vector in the radial direction.
This ``local capacitance approximation" is valid as long as $d$ is much smaller than any lateral length scale~\cite{dyakonov_prl_1993}. 
The electron velocity ${\bm v}(r,t)$, with radial and azimuthal components $v_{r}(r,t)$ and $v_{\theta}(r,t)$, respectively, fluctuates in response to the total electric field 
${\bm E}(r,t) = {\bm E}_{\hat{\bm \theta}}(r,t) + {\bm E}_{\hat{\bm r}}(r,t)$.
\begin{figure}[t]
\includegraphics[width=\columnwidth]{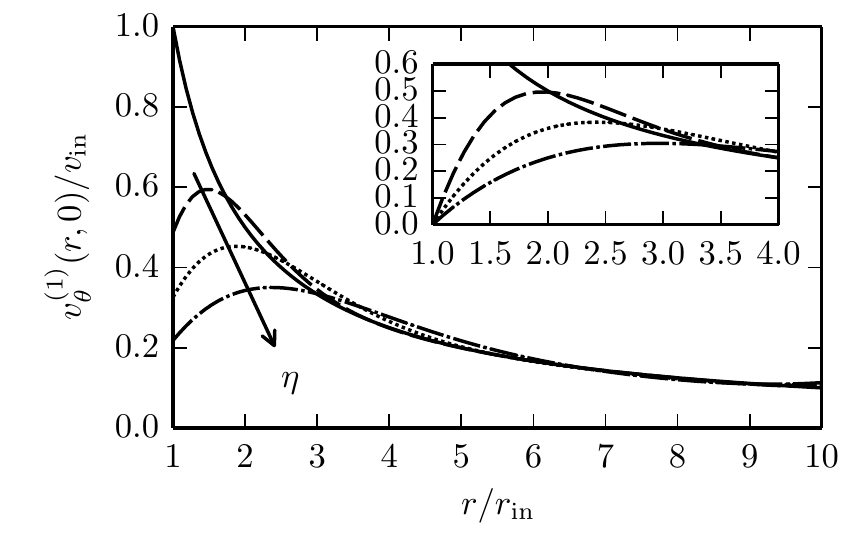}
\caption{\label{fig:profile}
Radial profile of the azimuthal component of the velocity---in units of $v_{\rm in}$ as defined in Eq.~(\ref{IrrotationalFlow})---at $t=0$, obtained by solving Eq.~(\ref{eq:profileeq}) with boundary conditions (\ref{eq:bcam}) (main panel) and (\ref{eq:bcvanish}) (inset).
The results are obtained with $r_{\rm out}/r_{\rm in} = 10.0$, and for several values of the dimensionless parameter $\xi$ defined in Eq.~(\ref{eq:dimlessparam}): $\xi = 0.2$ (dashed line), $0.5$ (dotted line), and $1.0$ (dash-dotted line).
The value of the viscosity in the different solutions increases as shown by the arrow.
The solid line corresponds to the analytical result~(\ref{IrrotationalFlow}), which holds at $\xi = 0$.}
\end{figure}

We now apply the Navier-Stokes equations of hydrodynamics~\cite{Landau06, batchelor} to the calculation of the time-evolution of $n(r,t)$ and ${\bm v}(r,t)$. 
At first order in $\Phi_{0}$, the electron density fluctuations $n^{(1)}(r,t)$ and the radial velocity fluctuations $v_{r}^{(1)}(r,t)$ vanish, while the azimuthal velocity oscillates at the frequency of the magnetic flux, $v_{\theta}^{(1)}(r,t) = \Re e[ v_{\theta}^{(1)}(r) e^{-i \Omega t}]$.
At second order in $\Phi_{0}$, the circular motion of the electrons in the CD orbiting around the $\hat{\bm z}$ axis generates a dc potential energy difference $\Delta U$ between the inner and outer edges of the CD, given by
\begin{equation}\label{eq:mainresult}
\Delta U = \frac{m}{2} \int_{r_{\rm in}}^{r_{\rm out}} dr \frac{1}{r} |v_{\theta}^{(1)}(r)|^{2}~,
\end{equation}
where $m$ is the effective electron mass.
The quantity $\Delta U$ in Eq.~(\ref{eq:mainresult})---independent of time---can be readily recognized as the work of the centripetal force acting on an electron which drifts from the inner to the outer rim of the CD and is subject to a constant acceleration in the radial direction given by $|v_{\theta}^{(1)}(r)|^{2} / (2 r)$.
For an ideal fluid ($\eta = 0$), it is easily shown (see discussion below) that the flow is given by
\begin{equation}\label{IrrotationalFlow}
\left.v_{\theta}^{(1)}(r) \right|_{\eta = 0}= \frac{e}{mc}\frac{\Phi_0}{2 \pi r} \equiv v_{\rm in} \frac{r_{\rm in}}{r}~,
\end{equation}
which is irrotational (i.e.~curl-free) in the region $r_{\rm in} < r < r_{\rm out}$. 
Putting this in Eq.~(\ref{eq:mainresult}) and assuming $r_{\rm out} \gg r_{\rm in}$ we obtain
\begin{equation}\label{eq:resnovisc}
\left. \Delta U  \right|_{\eta = 0} = \frac{e^{2}}{m c^{2}} \frac{1}{(2 \pi r_{\rm in})^{2}} \frac{1}{4} \Phi_{0}^{2}~.
\end{equation}
Notice that $\left. \Delta U  \right|_{\eta = 0}$ is independent of the frequency $\Omega$ of the driving flux.
To estimate the magnitude of the dc response, we use the following parameters~\cite{footnote_flux}:
\begin{eqnarray}\label{eq:paramexample}
& r_{\rm in} = 2.0~\mu{\rm m},\quad
r_{\rm out} = 20.0~\mu{\rm m}, \quad
& \nonumber \\
& \Phi_0 / (\pi r_{\rm in}^{2}) = 10~{\rm mT},\quad
m = 0.067~m_{\rm e}~,
\end{eqnarray}
where  $m_{\rm e}$ is the bare electron mass in vacuum and the value of $m$ given above is appropriate for electrons in GaAs~\cite{MBE}.
From Eq.~(\ref{eq:resnovisc}), we find $\Delta U |_{\eta = 0} = 63~\mu\rm{eV}$.
The analytical result (\ref{IrrotationalFlow}) is plotted as a solid line in Fig.~\ref{fig:profile}.

\begin{figure}[t]
\includegraphics[width=\columnwidth]{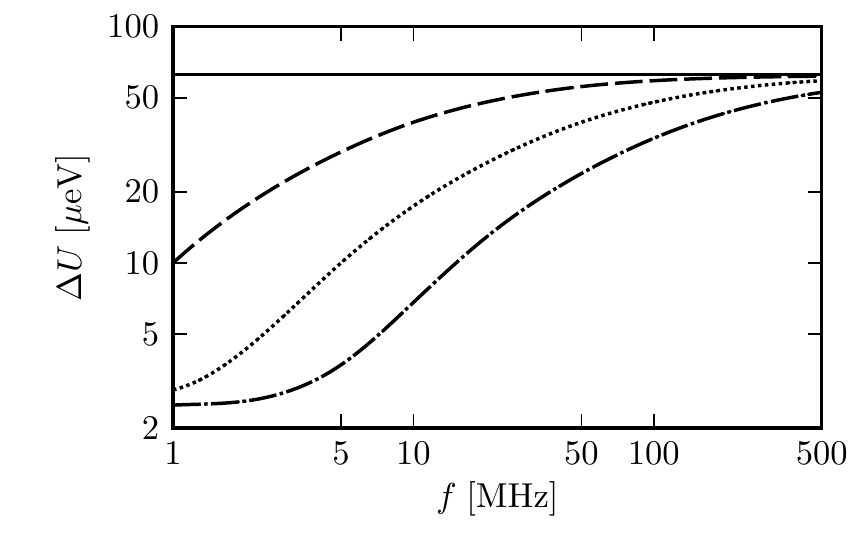}
\caption{The dc potential energy difference $\Delta U$ between outer and inner rims of the Corbino disk as a function of the frequency $f = \Omega / (2 \pi)$ of the oscillating magnetic flux $\Phi(t)$. Results in this plot have been obtained with the parameters as in Eq.~(\ref{eq:paramexample}). Different curves correspond to different values of the kinematic viscosity: $\nu = 1.0~{\rm cm}^{2}/{\rm s}$ (dashed line), $5.0~{\rm cm}^{2}/{\rm s}$ (dotted line), and $15.0~{\rm cm}^{2}/{\rm s}$ (dash-dotted line).
The solid line represents the viscous-free analytical result in Eq.~(\ref{eq:resnovisc}). \label{fig:response}}
\end{figure}

\begin{figure}[t]
\includegraphics[width=\columnwidth]{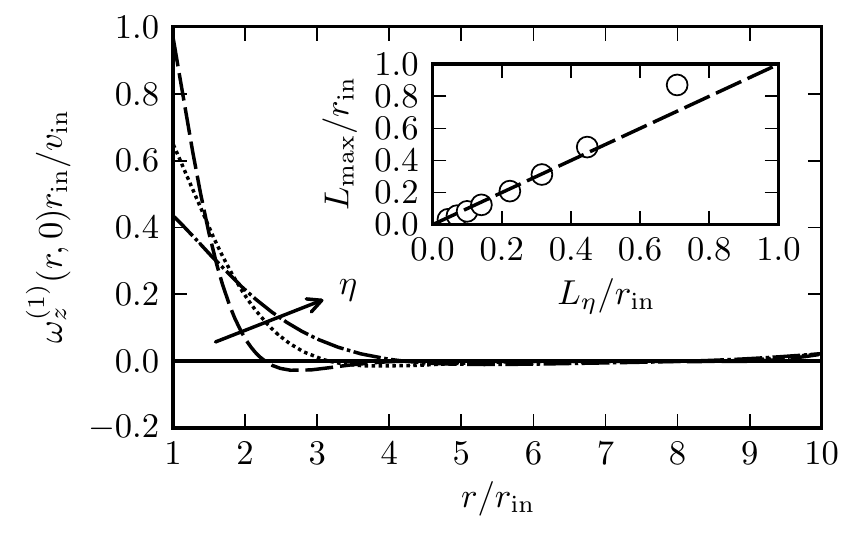}
\caption{Same as in Fig.~\ref{fig:profile} but for the radial profile of the vorticity $\omega^{(1)}_z(r,t)$ (in units of $v_{\rm in}/r_{\rm in}$) at time $t=0$. The inset shows the position $L_{\rm max}$ of the maximum of $v^{(1)}_\theta(r,t=0)$, as displayed in Fig.~\ref{fig:profile}, for several values of the vorticity penetration depth $L_\eta$ (in units of $r_{\rm in}$). The dashed line corresponds to the expected relation $L_{\rm max} = L_\eta$.\label{fig:vorticity}}
\end{figure}

Including the shear viscosity has three main effects on which we further elaborate below:  (i) the spatial variation of the velocity field is considerably reduced (see Fig.~\ref{fig:profile}), (ii) the flow acquires a non-curl-free dependence on the radial position $r$, i.e., a non-zero  {\it vorticity}~\cite{Landau06, batchelor} ${\bm \omega} = \nabla_{\bm r} \times {\bm v}$ appears near the inner and outer edges (see Fig.~\ref{fig:vorticity}), and (iii) the dc potential drop $\Delta U$ becomes strongly frequency-dependent (see Fig.~\ref{fig:response}).
Indeed, $\Delta U$ decreases by a factor $20$ as the frequency decreases from $500~{\rm MHz}$ (where the effect the viscosity is practically negligible) to $1~{\rm MHz}$. The profile of $\Delta U$ depends only on the ratio of the viscosity to the average mass density $m \bar{n}$, i.e.~the kinematic viscosity~\cite{Landau06, batchelor}
\begin{equation}\label{eq:kinematicviscosity}
\nu = \frac{\eta}{m \bar{n}}~.
\end{equation}
By measuring the frequency dependence of $\Delta U$, subtracting the frequency-independent background (\ref{eq:resnovisc}), and fitting the theoretical curve to the experimental result one can determine $\nu$. In the remainder of this Letter we supply the main steps of the calculation of $\Delta U$.

\noindent {\it Hydrodynamic equations and their solution.---}In the hydrodynamic regime, the response of the 2DQEL is governed by the Navier-Stokes equation~\cite{Landau06, batchelor}
\begin{equation}\label{eq:NavierStokes}
\begin{split}
& \rho(r,t) \left \lbrace \partial_{t}{\bm v}(r,t) + \left \lbrack {\bm v}(r,t) \cdot \nabla_{\bm r} \right \rbrack {\bm v}(r,t) \right \rbrace = \\
& -e {\bm E}(r,t) n(r,t) - \nabla_{\bm r} P(r,t) + \eta \nabla^2_{\bm r} {\bm v}(r,t) \\
& + \zeta \nabla_{\bm r} \left \lbrack \nabla_{\bm r} \cdot {\bm v}(r,t) \right \rbrack~,
\end{split}
\end{equation}
combined with the continuity equation
\begin{equation}\label{eq:continuity}
\partial_{t} n(r,t) + \nabla_{\bm r} \cdot \lbrack n(r,t) {\bm v}(r,t) \rbrack = 0~.
\end{equation}
Here, $\rho(r,t) = m n(r,t)$ is the mass density, $P(r,t)$ is the pressure, and $\zeta$ is the bulk viscosity, which we can neglect here since it vanishes at long wavelengths~\cite{Landau06, Giuliani_and_Vignale}.
In Eq.~(\ref{eq:NavierStokes}) we have also included the contribution of the electric field ${\bm E}(r,t)$. 
The azimuthal symmetry of the system implies that all quantities depend on the radial coordinate $r$ only and that the derivatives with respect to the azimuthal angle $\theta$ vanish. 
The radial component ${\bm E}_{\hat {\bm r}}(r,t)$ of the electric field is given by ${\bm E}_{\hat {\bm r}}(r,t) = - \hat {\bm r} \partial_r U(r,t)/e$ where the 
electric potential energy $U(r,t)$ is obtained by solving the Poisson equation in the CD geometry with a constant boundary condition at the gate position $z = -d$.
If the typical wavelength of density fluctuations is larger than $d$, it is easy to see that
\begin{equation}\label{eq:gca}
U(r,t) \simeq -e^{2} n(r,t) / C~,
\end{equation}
which immediately leads to Eq.~(\ref{eq:Eradial}). Finally, the pressure gradient in Eq.~(\ref{eq:NavierStokes}) can be neglected when $\partial P(n)/\partial n \ll e^2 \bar{n}/C$, i.e. when $d \gg a^\star_B/4$. This inequality is always well satisfied since $a^\star_{\rm B} \equiv \epsilon \hbar^2/(m e^2)$ is the material Bohr radius, which is $\sim 10~{\rm nm}$ for GaAs.

The Navier-Stokes and continuity equations (\ref{eq:NavierStokes})-(\ref{eq:continuity}) must be complemented by suitable boundary conditions (BCs) expressed in terms of the flux density tensor~\cite{Landau06, batchelor} $\Pi_{i,k}(r,t) = P(r,t) \delta_{i,k} + \rho v_{i}(r,t)v_{k}(r,t) - \sigma_{i,k}'(r,t)$,
where $\sigma_{i,k}'(r,t)$ is the viscous stress tensor.
We require the radial diffusion of azimuthal momentum, which is proportional to the viscosity $\eta$, to vanish at the outer and inner rims of the CD, i.e.~
\begin{equation}\label{eq:boundary}
\left.\sigma_{r,\theta}'(r, t)\right|_{r= r_{\rm in}} = \left.\sigma_{r,\theta}'(r, t)\right|_{r= r_{\rm out}} \equiv 0~.
\end{equation}
The off-diagonal component of the viscous stress tensor reads $\sigma_{r,\theta}' = \eta ( \partial_{r} v_{\theta} + \partial_{\theta} v_{r} / r - v_{\theta} / r)$~\cite{Landau06, batchelor}.
Because of circular symmetry, the BCs (\ref{eq:boundary}) reduce to
\begin{equation}\label{eq:bcam}
\partial_{r} v_{\theta}(r,t) |_{r = r_{i}} = \left.\frac{v_{\theta}(r,t)}{r}\right|_{r = r_{i}}~,
\end{equation}
where $r_{i} = r_{\rm in}, r_{\rm out}$.
Moreover, for the setup in Fig.~\ref{fig:setup}, two further BCs should be imposed.
First, the radial component of the current ${\bm j}(r,t) = n(r,t) {\bm v}(r,t)$ must vanish at the outer rim, where the CD is isolated.
Second, the electric potential at the inner rim is fixed, hence the electron density
$n(r_{\rm in},t)$ at the inner rim of the CD must coincide with the average value $\bar{n} = e V_{\rm G} / C$ fixed by the gate voltage $V_{\rm G}$.

We notice that the BCs~(\ref{eq:bcam}) differ from the standard ``no-slip'' BCs
\begin{equation}\label{eq:bcvanish}
v_{\theta}(r,t) |_{r = r_{\rm in}} = v_{\theta}(r,t) |_{r = r_{\rm out}} \equiv 0~,
\end{equation}
which are commonly employed~\cite{Landau06, batchelor} to describe fluid adhesion to the walls of a container.
While the use of the BCs in Eq.~(\ref{eq:bcvanish}) is not immediately justified in our case, we have checked that the results in Fig.~\ref{fig:response} do not change qualitatively if the BCs in Eq.~(\ref{eq:bcvanish}) are used instead of those in Eq.~(\ref{eq:bcam}).
The agreement between the results obtained with two different sets of BCs gives us confidence in the robustness of the effect illustrated in Fig.~\ref{fig:response}.

We solve Eqs.~(\ref{eq:NavierStokes})-(\ref{eq:continuity}) by expanding the hydrodynamic variables in powers of the amplitude $\Phi_{0}$ of the magnetic flux~\cite{dyakonov_ieee_1996a}:
\begin{align}\label{eq:expansion}
{\bm v}(r, t) & = {\bm v}^{(0)}(r, t) + {\bm v}^{(1)}(r, t) + [ \delta {\bm v}(r) + {\bm v}^{(2)}(r, t) ] + \dots \nonumber \\
n(r,t) & = n^{(0)}(r, t) + n^{(1)}(r, t) + [ \delta n(r) + n^{(2)}(r, t) ] + \dots,
\end{align}
where ${\bm v}^{(k)}(r,t),n^{(k)}(r,t) \sim (\Phi_{0})^k \cos{(k \Omega t)}$.
Since the hydrodynamic equations of motion are nonlinear, in the expansion~(\ref{eq:expansion}) we include constant contributions $\delta{\bm v}(r),\delta n(r) \sim {\cal O}(\Phi^2_0)$ arising from the self-mixing of the signal at frequency $\Omega$~\cite{dyakonov_ieee_1996a}.
At first order, we find the following differential equation for the radial profile of the azimuthal component of the velocity:
\begin{eqnarray}\label{eq:profileeq}
-i \Omega v_{\theta}^{(1)}(r) & =&
\nu \left \lbrack \partial_{r}^{2} v_{\theta}^{(1)}(r) + \frac{1}{r} \partial_{r} v_{\theta}^{(1)}(r) - \frac{1}{r^{2}} v_{\theta}^{(1)}(r) \right \rbrack \nonumber\\ 
& -& i \Omega \frac{e}{m c} \frac{\Phi_{0}}{2 \pi r}~.
\end{eqnarray}
In the general $\nu \neq 0$ case, Eq.~(\ref{eq:profileeq}) must be solved numerically~\cite{footnotenumerics}. It is convenient to introduce dimensionless variables by rescaling $r$ 
with $r_{\rm in}$ and $v_{\theta}^{(1)}(r)$ with $v_{\rm in}$, which has been defined in Eq.~(\ref{IrrotationalFlow}). Then, Eq.~(\ref{eq:profileeq}) depends only on the dimensionless parameter
\begin{equation}\label{eq:dimlessparam}
\xi \equiv \left(\frac{L_\eta}{r_{\rm in}}\right)^2~,
\end{equation}
where $L_\eta = \sqrt{\nu/\Omega}$ is the {\it vorticity penetration depth}~\cite{Landau06, batchelor} during a time $1/\Omega$.  The solution of Eq.~(\ref{eq:profileeq}) is shown in Fig.~\ref{fig:profile} for several values of $\xi$ and both sets of BCs. We note that with increasing viscosity the amplitude of the velocity flow diminishes. Since according to Eq.~(\ref{eq:mainresult}) $\Delta U$ depends on the integral of the square of the velocity profile, increasing the viscosity suppresses the dc response $\Delta U$, thereby explaining the results in Fig.~\ref{fig:response}.

In the limit $\xi \ll 1$ (low viscosity or high frequency) the solution of Eq.~(\ref{eq:profileeq}) is given by the curl-free profile in Eq.~(\ref{IrrotationalFlow}).  In the opposite $\xi \gg 1$ limit (high viscosity or low frequency) the solution is found by setting to zero the term in square brackets in Eq.~(\ref{eq:profileeq}). In this case, it is easy to demonstrate that a non-curl-free linear profile $v_{\theta}^{(1)}(r) \propto r$ solves the problem, satisfying the BCs. 

To obtain the dc~response $\Delta U$ in Eq.~(\ref{eq:mainresult}) we  expand Eq.~(\ref{eq:continuity}) and the radial component of Eq.~(\ref{eq:NavierStokes}) to second order and we average the resulting expressions over a period $T = 2\pi / \Omega$ of the oscillating flux $\Phi(t)$.
The solution of the time-averaged equations yields $\delta v_{r}(r) \equiv 0$ and
\begin{equation}\label{eq:secondorder}
\delta n(r) = \frac{m C}{e^{2}} \int_{r_{\rm in}}^{r} dr' \frac{1}{r'} \langle v_{\theta}^{(1)}(r,t)^{2} \rangle_{t}~,
\end{equation}
where $\langle g(t) \rangle_{t} \equiv T^{-1}\int_{0}^{T}dt' g(t')$ denotes the time-average of a function $g(t)$ over one period of the oscillating magnetic flux.
Finally, Eq.~(\ref{eq:mainresult}) can be easily obtained by setting $r = r_{\rm out}$ in Eq.~(\ref{eq:secondorder}) and making use of $\langle v_{\theta}^{(1)}(r,t)^{2} \rangle_{t} = |v_{\theta}^{(1)}(r)|^{2} / 2$ and  of the local capacitance formula, Eq.~(\ref{eq:gca}).

\noindent {\it Vorticity and dissipation.---}Further insights on the physical properties of the solution shown in Fig.~\ref{fig:profile} can be obtained by looking at the radial profile of the vorticity 
$\omega_z(r,t) = \partial_r [r v_\theta(r,t)]/r$, whose first-order contribution $\omega^{(1)}_{z}(r,t)$ in powers of $\Phi_0$ is shown in Fig.~\ref{fig:vorticity}. In the regions near the rims of the CD the non-curl-free dependence of the velocity flow $v^{(1)}_\theta(r,t)$ on $r$ leads to large values of the vorticity. Moving away from the rims, the vorticity decreases to zero on a length scale $L_\eta$, which confirms the interpretation of this quantity as the vorticity penetration depth~\cite{Landau06,batchelor,footnote_nu}. On the same length scale the velocity flow crosses over to the curl-free profile (\ref{IrrotationalFlow})---see inset in Fig.~\ref{fig:vorticity}.

Before concluding, we would like to quantify the power $W$ dissipated in the fluid by viscosity~\cite{Landau06, batchelor} in a period of the oscillating flux. 
This is given by the following expression:
\begin{equation}\label{eq:instpower}
W = - \pi \eta \int_{r_{\rm in}}^{r_{\rm out}} dr~\frac{1}{r}~\left| v_{\theta}^{(1)}(r) - r\partial_r v_{\theta}^{(1)}(r) \right|^{2}~.
\end{equation}
To leading order in $\eta$ for $\eta \to 0$ we obtain $W \to  - 8 \pi (\eta/m) \Delta U|_{\eta = 0}$.  As far as order-of-magnitudes are concerned, we find that $W$ is of the order of microwatts for $\nu \sim 10~{\rm cm}^{2}/s$ and ${\bar n} \sim 10^{11}~{\rm cm}^{-2}$. Away from the perturbative regime, we have verified that $\eta \sim - a m W/(8\pi \Delta U)$ where $a \sim 0.5$ is a numerical coefficient. This relation is valid in the range $0.25 \lesssim \xi \lesssim 1.0$.  Thus, an independent measure of $W$ {\it and} $\Delta U$ {\it at a single frequency $\Omega$} of the oscillating magnetic flux yields the value of $\eta$. This provides a possible alternative to the method described after Eq.~(\ref{eq:kinematicviscosity}), which requires a measurement of $\Delta U$ as a function of frequency.

In summary, we have demonstrated that the hydrodynamic shear viscosity of a two-dimensional quantum electron liquid can be obtained by studying the response of the system to an oscillating magnetic flux in a Corbino disk geometry. We truly hope that this work will stimulate further studies of viscometers for two-dimensional quantum electron liquids and related experimental activities on the shear viscosity of these systems, which may pave the way for the discovery of solid-state nearly perfect fluids~\cite{NPFsreview}.

\acknowledgments
G.V.~was supported by the NSF through Grant No. DMR-1104788.
We gratefully acknowledge A.~Hamilton for very useful discussions.
We have made use of free software (www.gnu.org, www.python.org).


\begin{thebibliography}{99}

\bibitem{Landau06}
L.D. Landau and E.M. Lifshitz, {\it Course of Theoretical Physics: Fluid Mechanics} (Pergamon, New York, 1987).

\bibitem{batchelor}
G.K. Batchelor, {\it An Introduction to Fluid Dynamics} (Cambridge University Press, Cambridge, 1967).

\bibitem{maxwell}
J.C. Maxwell, \href{http://www.jstor.org/stable/108948}{Philos. Trans. R. Soc. London {\bf 156}, 249 (1866)}.

\bibitem{abrikosovkhalatnikov}
A.A. Abrikosov and I.M. Khalatnikov, \href{http://dx.doi.org/10.1088/0034-4885/22/1/310}{Rep. Prog. Phys. {\bf 22}, 329 (1959)}.

\bibitem{NPFsreview}
T. Sch\"{a}fer and D. Teaney, \href{http://dx.doi.org/10.1088/0034-4885/72/12/126001}{Rep. Prog. Phys. {\bf 72}, 126001 (2009)}.

\bibitem{cao_science_2011}
C. Cao, E. Elliott, J. Joseph, H. Wu, J. Petricka, T. Sch\"{a}fer, and J.E. Thomas, \href{http://dx.doi.org/10.1126/science.1195219}{Science {\bf 331}, 58 (2011)}.

\bibitem{cao_njp_2011}
C. Cao, E. Elliott, H. Wu, and J.E. Thomas, \href{http://dx.doi.org/10.1088/1367-2630/13/7/075007}{New J. Phys. {\bf 13}, 075007 (2011)}.

\bibitem{vogt_prl_2012}
E. Vogt, M. Feld, B. Fr\"{o}hlich, D. Pertot, M. Koschorreck, and M. K\"{o}hl, \href{http://dx.doi.org/10.1103/PhysRevLett.108.070404}{Phys. Rev. Lett. {\bf 108}, 070404 (2012)}.

\bibitem{QGP}
R. Snellings, \href{http://dx.doi.org/10.1088/1367-2630/13/5/055008}{New J. Phys. {\bf 13}, 055008 (2011)}.
B.V. Jacak and B. M\"{u}ller, \href{http://dx.doi.org/10.1126/science.1215901}{Science {\bf 337}, 310 (2012)}.

\bibitem{MBE}
L.N. Pfeiffer and K.W. West, \href{http://dx.doi.org/10.1016/j.physe.2003.09.035}{Physica E {\bf 20}, 57 (2003)};
V. Umansky, M. Heiblum, Y. Levinson, J. Smet, J. N\"{u}bler, and M. Dolev, \href{http://dx.doi.org/10.1016/j.jcrysgro.2008.09.151}{J. Cryst. Growth {\bf 311}, 1658 (2009)}.

\bibitem{other2DEGs}
For 2DQELs other than 2D electron gases in semiconductor heterostructures see, for example,  K.S. Novoselov, D. Jiang, F. Schedin, T.J. Booth, V.V. Khotkevich, S.V. Morozov, and A.K. Geim, \href{http://dx.doi.org/10.1073/pnas.0502848102}{Proc. Natl. Acad. Sci. (USA) {\bf 102}, 10451 (2005)};
K.S. Novoselov and A.K. Geim, \href{http://dx.doi.org/10.1038/nmat1849}{Nature Mater. {\bf 6}, 183 (2007)};
M.Z. Hasan and C.L. Kane, \href{http://dx.doi.org/10.1103/RevModPhys.82.3045}{Rev. Mod. Phys. {\bf 82}, 3045 (2010)};
X.-L. Qi and S.-C. Zhang, \href{http://dx.doi.org/10.1103/RevModPhys.83.1057}{{\it ibid.} {\bf 83}, 1057 (2011)};
J. Mannhart,  D.H.A. Blank, H.Y. Hwang, A.J. Millis, and J.-M. Triscone, \href{http://dx.doi.org/10.1557/mrs2008.222}{MRS Bullettin {\bf 33}, 1027 (2008)}; J. Mannhart and D.G. Schlom, \href{http://dx.doi.org/10.1126/science.1181862}{Science {\bf 327}, 1607 (2010)} and references therein.

\bibitem{shear_viscosity_2DQELs}
See, for example, S. Conti and G. Vignale, \href{http://dx.doi.org/10.1103/PhysRevB.60.7966}{Phys. Rev. B {\bf 60}, 7966 (1999)}; I.V. Tokatly and G. Vignale, \href{http://dx.doi.org/10.1103/PhysRevB.76.161305}{{\it ibid.}~{\bf 76}, 161305(R) (2007)}; \href{http://dx.doi.org/10.1103/PhysRevB.79.199903}{{\bf 79}, 199903(E) (2009)}; F.D.M. Haldane, \href{http://arxiv.org/abs/0906.1854}{arXiv:0906.1854 (2009)}; N. Read, \href{http://dx.doi.org/10.1103/PhysRevB.79.045308}{Phys. Rev. B {\bf 79}, 045308 (2009)}; M. M\"uller, J. Schmalian, and L. Fritz, \href{http://dx.doi.org/10.1103/PhysRevLett.103.025301}{Phys. Rev. Lett. {\bf 103}, 025301 (2009)}.
T.L. Hughes, R.G. Leigh, and E. Fradkin, \href{http://dx.doi.org/10.1103/PhysRevLett.107.075502}{{\it ibid.} {\bf 107}, 075502 (2011)}; B. Bradlyn, M. Goldstein, and N. Read, \href{http://dx.doi.org/10.1103/PhysRevB.86.245309}{Phys. Rev. B {\bf 86}, 245309 (2012)}; C. Hoyos and D.T. Son, \href{10.1103/PhysRevLett.108.066805}{Phys. Rev. Lett. {\bf 108}, 066805 (2012)} and references therein.

\bibitem{Giuliani_and_Vignale}
G.F. Giuliani and G. Vignale, {\it Quantum Theory of the Electron Liquid} (Cambridge University Press, Cambridge, 2005).

\bibitem{dyakonov_prl_1993}
M. Dyakonov and M. Shur, \href{http://dx.doi.org/10.1103/PhysRevLett.71.2465}{Phys. Rev. Lett. {\bf 71}, 2465 (1993)}.

\bibitem{govorov_prl_2004}
A.O. Govorov and J.J. Heremans, \href{http://dx.doi.org/10.1103/PhysRevLett.92.026803}{Phys. Rev. Lett. {\bf 92}, 026803 (2004)}.

\bibitem{andreev_prl_2011}
A.V. Andreev, S.A. Kivelson, and B. Spivak, \href{http://dx.doi.org/10.1103/PhysRevLett.106.256804}{Phys. Rev. Lett. {\bf 106}, 256804 (2011)}.

\bibitem{hydrodynamicsexperimental}
See, for example, M.J.M. de Jong and L.W. Molenkamp, \href{http://dx.doi.org/10.1103/PhysRevB.51.13389} {Phys. Rev. B  {\bf 51}, 13389 (1995)}; D. Taubert, G.J. Schinner, H.P. Tranitz, W. Wegscheider, C. Tomaras, S. Kehrein, and S. Ludwig, \href{http://dx.doi.org/10.1103/PhysRevB.82.161416}{{\it ibid.}~{\bf 82}, 161416 (2010)}; D. Taubert, G.J. Schinner, C. Tomaras, H.P. Tranitz, W. Wegscheider, and S. Ludwig, \href{http://dx.doi.org/10.1063/1.3577959}{J. Appl. Phys. {\bf 109}, 102412 (2011)}.

\bibitem{footnote_flux}
The value of $\Phi_0$ given in Eq.~(\ref{eq:paramexample}) corresponds $\Phi_{0} \simeq 60~\phi_{0}$, where $\phi_{0} = h / (2 e) \simeq 2\times 10^{-15}~{\rm Wb}$ is the magnetic flux quantum.

\bibitem{dyakonov_ieee_1996a}
M. Dyakonov and M. Shur, \href{http://dx.doi.org/10.1109/16.485650}{IEEE Trans. Electron Devices {\bf 43}, 380 (1996)}.

\bibitem{footnotenumerics}
It is numerically convenient to solve Eq.~(\ref{eq:profileeq}) by adding a friction term $-\gamma v_{\theta}^{(1)}(r)$ with $\gamma = 10^{-3}\Omega$ to the right-hand side.

\bibitem{footnote_nu}
This can be seen as follows: $\omega^{(1)}_{z}(r,t)$ obeys the 2D equation of motion~\cite{batchelor} $\partial_{t} \omega^{(1)}_{z}(r,t) = \nu \nabla^{2}_{\bm r} \omega^{(1)}_{z}(r,t)$. Therefore, $\nu$ plays the role of diffusion constant for $\omega^{(1)}_{z}(r,t)$. This happens because the 2D flow in our case is uniform and incompressible to ${\cal O}(\Phi_{0})$.

\end{thebibliography}
\end{document}